%% file: main.tex
\documentclass[aps,twocolumn,raggedbottom,superscriptaddress,tightenlines,pra,10pt,longbibliography, nofootinbib, floatfix]{revtex4-1}
\usepackage[utf8]{inputenc}
\usepackage{amsthm}
\usepackage{amsmath}
\usepackage{amssymb}
\usepackage{mathtools}
\usepackage{graphicx}
\usepackage[font={small}]{caption}
\usepackage{subcaption}
\usepackage{epstopdf}
\usepackage{url}
\usepackage{float}
\usepackage{bm}
\usepackage{ifthen}
\usepackage[usenames,dvipsnames]{color}
\usepackage{mathrsfs}
\usepackage{blkarray}
\usepackage{natbib}
\usepackage[colorlinks=true,citecolor=blue,urlcolor=blue,linkcolor=blue]{hyperref}
\usepackage{epstopdf}
\usepackage{braket}
\usepackage{array}
\usepackage{soul}
\usepackage[english]{babel}
\usepackage{algorithm}
\usepackage{algpseudocode}
\usepackage{textcomp}
\usepackage{xcolor}
\usepackage{xspace}
\normalsize
\usepackage{booktabs}
\usepackage{multirow}

\usepackage{lineno}

\newcommand{\ignore}[1]{}

\DeclareMathOperator{\inp}{\text{in}}
\DeclareMathOperator{\out}{\text{out}}
\DeclareMathOperator{\adc}{\text{ADC}}
\DeclareMathOperator{\dac}{\text{DAC}}

\bibpunct{}{}{,}{s}{}{\textsuperscript{,}}

\begin{document}
\title{A blueprint for precise and fault-tolerant analog neural networks}

\author{Cansu Demirkiran}
\email[E-mail: ]{cansu@bu.edu}
\affiliation{Boston University}
\author{Lakshmi Nair}
\affiliation{Lightmatter}
\author{Darius Bunandar}
\affiliation{Lightmatter}
\author{Ajay Joshi}
\affiliation{Boston University}


\input{abstract}
\input{text/1--introduction}

\input{text/2--results}
\input{text/3--discussion}
\input{text/4--methods}

\input{acknowledgements}
\input{contributions}
\input{competing_interests}
\newpage
\input{references}


\newpage
\end{document}

%% file: abstract.tex

\onecolumngrid
\maketitle
\textbf{Abstract}

Analog computing has reemerged as a promising avenue for accelerating deep neural networks (DNNs) due to its potential to overcome the energy efficiency and scalability challenges posed by traditional digital architectures.
However, achieving high precision and DNN accuracy using such technologies is challenging, as high-precision data converters are costly and impractical.
In this paper, we address this challenge by using the residue number system (RNS).
RNS allows composing high-precision operations from multiple low-precision operations, thereby eliminating the information loss caused by the limited precision of the data converters.
Our study demonstrates that analog accelerators utilizing the RNS-based approach can achieve ${\geq}99\%$ of FP32 accuracy for state-of-the-art DNN inference using data converters with only $6$-bit precision whereas a conventional analog core requires more than $8$-bit precision to achieve the same accuracy in the same DNNs.
The reduced precision requirements imply that using RNS can reduce the energy consumption of analog accelerators by several orders of magnitude while maintaining the same throughput and precision.
Our study extends this approach to DNN training, where we can efficiently train DNNs using $7$-bit integer arithmetic while achieving accuracy comparable to FP32 precision.
Lastly, we present a fault-tolerant dataflow using redundant RNS error-correcting codes to protect the computation against noise and errors inherent within an analog accelerator.

\vspace{0.2in}

%% file: text/1--introduction.tex
\twocolumngrid

\section*{Introduction}
\label{sec:introduction}

Deep Neural Networks (DNNs) are widely employed across various applications today.
Unfortunately, their compute, memory, and communication demands are continuously on the rise.
The slow-down in CMOS technology scaling, along with these increasing demands has led analog DNN accelerators to gain significant research interest. 
Recent research has been focused on using various analog technologies such as photonic cores~\cite{CoherentNanophotonic2017, 11-tops, weight-bank-2017, dnnara, pixel-2020, albireo-2021, adept}, resistive arrays~\cite{yao2020fully, prime-2016, isaac, 1t1m-2016, tang-2017}, switched capacitor arrays~\cite{bankman-2015, bankman-sc16}, Phase Change Materials (PCM)~\cite{feldmann2021parallel}, Spin-Transfer Torque (STT)-RAM~\cite{jain2017computing, shi2020performance}, etc., to enable highly parallel, fast, and efficient matrix-vector multiplications (MVMs) in the analog domain. 
These MVMs are fundamental components used to build larger general matrix-matrix multiplication (GEMM) operations, which make up more than $90\%$ of the operations in DNN inference and training~\cite{chen2020survey}.

The success of this approach, however, is constrained by the limited precision of the digital-to-analog and analog-to-digital data converters (i.e., DACs and ADCs).
In an analog accelerator, the data is converted between analog and digital domains using data converters before and after every analog operation. 
Typically, a complete GEMM operation cannot be performed at once in the analog domain due to the fixed size of the analog core. Instead, the GEMM operation is tiled into smaller MVM operations.  
As a result, each MVM operation produces a partial output that must be accumulated with other partial outputs to obtain the final GEMM result.
Concretely, an MVM operation consists of parallel dot products between $b_w$-bit signed weight vectors and $b_{in}$-bit signed input vectors---each with $h$ elements---resulting in a partial output containing $b_{\out}$ bits of information, where $b_{\out} = b_{\inp} + b_w + \log_2(h)-1$. 
An ADC with a precision greater than $b_{\out}$ (i.e., $b_{\adc} \geq b_{\out}$) is required to ensure no loss of information when capturing these partial outputs.
Unfortunately, the energy consumption of ADCs increases exponentially with bit precision (often referred to as effective number of bits (ENOB)). 
This increase is roughly 4$\times$ for each additional output bit~\cite{murmann-mixed-signal}.

\begin{figure}[t]
\centering
\includegraphics[width=\linewidth]{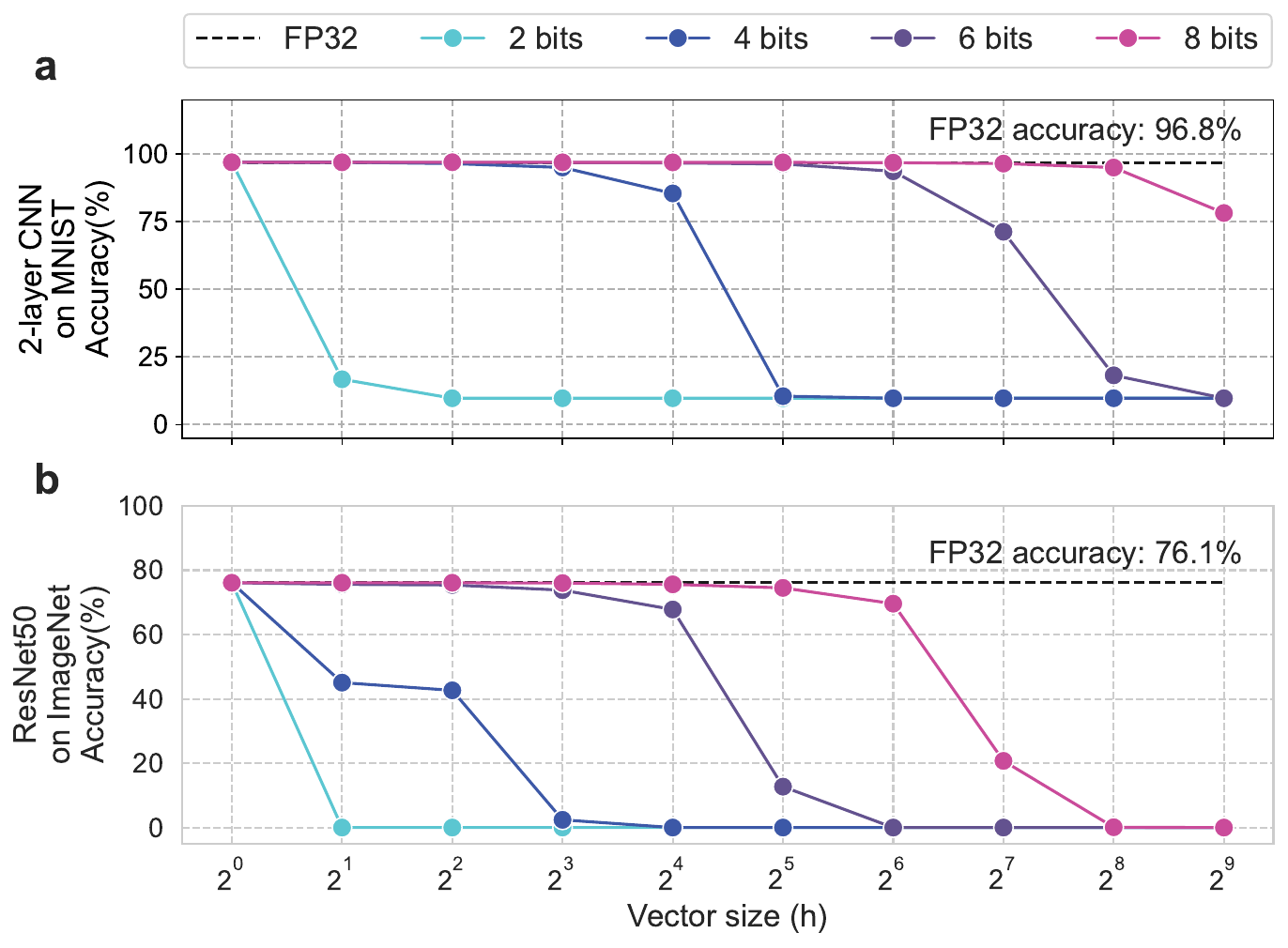}
\caption{\scriptsize{\textbf{Inference accuracy versus vector size for varying data bit-width in a conventional analog core.} \textbf{a} Inference accuracy for a two-layer CNN classifying handwritten digits from the MNIST dataset. \textbf{b} Inference accuracy for ResNet50 classifying images from the ImageNet dataset evaluated in an analog core with varying precision $b$ and vector sizes $h$. For both \textbf{a} and \textbf{b}, $b$-bit precision means $b =b_{\dac}=b_{\adc} < b_\text{out}$ where $b$ varies between $2$ and $8$.}}
\label{fig:acc-tile-size}
\end{figure}

As a result, energy-efficient analog accelerator designs typically employ ADCs with lower precision than $b_{\out}$ and only capture the $b_{\adc}$ most significant bits (MSBs) from the $b_{\out}$ bits of each partial output~\cite{rekhi-2019}.
Reading only MSBs causes information loss in each partial output leading to accuracy degradation in DNNs, as pointed out by Rekhi et al.~\cite{rekhi-2019}. 
This degradation is most pronounced in large DNNs and large datasets. 
Fig.~\ref{fig:acc-tile-size} shows the impact of this approach on DNN accuracy in two tasks: (1) a two-layer convolutional neural network (CNN) for classifying the MNIST dataset~\cite{mnist}: a simple task with only 10 classes, and (2) the ResNet50 CNN~\cite{resnet} for classifying the ImageNet dataset~\cite{imagenet}: a more challenging task with 1000 classes.
As the vector size $h$ increases, higher precision is needed at the output to maintain the accuracy in both DNNs.
Moreover, ResNet50 experiences accuracy degradation at smaller values of $h$ compared to the two-layer CNN. 
While using a higher precision ADC can help recover from this accuracy degradation, it significantly reduces the energy efficiency of the analog hardware. 
Essentially, to efficiently execute large DNNs using analog accelerators, it is crucial to find a better way to achieve high accuracy than simply increasing the bit precision of the data converters.

In this work, we present a universal residue number system (RNS)-based framework to overcome the abovementioned challenge in analog DNN inference as well as DNN training.
RNS represents high-precision values using multiple low-precision integer residues for a selected set of moduli. 
As such, RNS enables high-precision arithmetic without any information loss on the partial products, even when using low-precision DACs and ADCs.
Utilization of RNS leads to a significant reduction in the data converter energy consumption, which is the primary contributor to energy usage in analog accelerators. 
This reduction can reach up to six orders of magnitude compared to a conventional fixed-point analog core with the same output bit precision. 

Our study shows that the RNS-based approach enables $\geq 99\%$ FP-32 inference accuracy by using only 6-bit data converters for state-of-the-art \textsf{MLPerf (Inference: Datacenters)} benchmarks~\cite{ml-perf-2020} and Large Language Models (LLMs). 
We also demonstrate the applicability of this approach in training and fine-tuning state-of-the-art DNNs using low-precision analog hardware.
The RNS approach, however, is susceptible to noise as small errors in the residues scale up during output reconstruction, leading to larger errors in the standard representation.
To address this issue, we incorporate the Redundant RNS (RRNS) error-correcting code~\cite{rrns-2015, rrns-codes-hanzo, yang2001redundant} to introduce fault-tolerance capabilities into the dataflow.

As RNS is closed under multiplication and addition, no significant changes are required in the design of the analog core or in how  GEMM operations are performed. 
Unlike a conventional analog core design, performing RNS operations necessitates an analog modulo operation.
This operation can be implemented by using ring oscillators~\cite{analog-modulo-2018} in an analog electrical core or by using optical phase shifters in an analog optical core. 
Our proposed framework, however, remains agnostic to the underlying technology. 
Importantly, arbitrary fixed-point precision can be achieved by combining the positional number system (PNS) and RNS in analog hardware.
Overall, our presented RNS-based methodology offers a solution combining high accuracy, high energy efficiency, and fault tolerance in analog DNN inference and training.

%% file: text/2--results.tex
\section*{Results}
\input{text/2-1-results-dnn-with-rns}

\input{text/2-2-results-precision}
\input{text/2-3-results-accuracy}
\input{text/2-4-results-error-corr}

%% file: text/2-1-results-dnn-with-rns.tex
\subsection*{DNN Inference and Training Using RNS}
\label{sec:rns-inf-tr}

The RNS represents an integer as a set of smaller (integer) residues.
These residues are calculated by performing a modulo operation on the said integer using a selected set of $n$ \emph{co-prime} moduli. 
Let $A$ be an integer. 
$A$ can be represented in the RNS with $n$ residues as $\{a_1, \dots, a_{n}\}$ for a set of co-prime moduli $\mathcal{M} = \{m_1, \dots, m_{n}\}$ where $a_i = |A|_{m_i} \equiv A \mod m_i$ for $i \in \{1\dots n\}$. 
$A$ can be uniquely reconstructed using the Chinese Remainder Theorem (CRT):
\begin{equation}
A = \bigg|\sum_{i=1}^na_i M_i T_i\bigg|_M,
    \label{eq:crt}
\end{equation}
if $A$ is within the range $[0, M)$ where $M= \prod_i m_i$. 
Here, $M_i = M/m_i$ and $T_i$ is the multiplicative inverse of $M_i$, i.e., $\left| M_i T_i \right|_{m_i} \equiv 1$. 
Hereinafter, we refer to the integer $A$ as the \emph{standard representation}, while we refer to the set of integers $\{ a_1, \dots, a_n\}$ simply as the residues. 

A DNN consists of a sequence of $L$ layers.
During inference, where the DNN is previously trained and its parameters are fixed, only a forward pass is performed.
Generically, the input $X$ to $({\ell}+1)$-th layer of a DNN during the forward pass is the output generated by the previous $\ell$-th layer:
\begin{equation}
    X^{(\ell+1)} = f^{(\ell)} \big( W^{(\ell)} X^{(\ell)} \big),
    \label{eq:nn}
\end{equation}
where $W^{(\ell)} X^{(\ell)}  = O^{(\ell)}$ is a GEMM operation and $f(\cdot)$ is an element-wise nonlinear function.

DNN training requires both forward and backward passes as well as weight updates. The forward pass in the training is performed the same way as in Eq.~\eqref{eq:nn}. 
After the forward pass, a loss value $\mathcal{L}$ is calculated using the output of the last layer and the ground truth. 
The gradients of the DNN activations and parameters with respect to $\mathcal{L}$ for each layer are calculated by performing a backward pass after each forward pass:

\begin{equation}
    \frac{\partial{\mathcal{L}}}{\partial {X^{(\ell)}}} =  {{W^{(\ell)}}^T} \frac{\partial{\mathcal{L}}}{\partial {O^{(\ell)}}},
    \label{eq:back-x}
\end{equation}
\begin{equation}
    \frac{\partial{\mathcal{L}}}{\partial {W^{(\ell)}}} =  \frac{\partial{\mathcal{L}}}{\partial {O^{(\ell)}}} {X^{(\ell)}}^T.
    \label{eq:back-w}
\end{equation}

Using these gradients $\Delta W^{(\ell)} = {\frac{\partial{\mathcal{L}}}{\partial {W^{(\ell)}}}}$, the DNN parameters are updated in each iteration $i$:
\begin{equation}
W^{(\ell)}_{i+1} = W^{(\ell)}_i - \eta \Delta W^{(\ell)}_i
    \label{eq:update-w}
\end{equation}
with a step size $\eta$ for a simple stochastic gradient descent (SGD) optimization algorithm. 

Essentially, for each layer, one GEMM operation is performed in the forward pass and two GEMM operations are performed in the backward pass. 
Because RNS is closed under addition and multiplication operations, GEMM operations can be performed in the RNS space. 
Using the RNS, Eq.~\eqref{eq:nn} can be rewritten as:
\begin{equation}
    X^{(\ell+1)} = f^{(\ell)} \Bigg(\text{CRT}\bigg(\Big|
    \big|W^{(\ell)}\big|_{\mathcal{M}}
    \big|{X}^{(\ell)}\big|_{\mathcal{M}}
     \Big|_{\mathcal{M}}\bigg)\Bigg).
    \label{eq:nn-rns}
\end{equation}
The same approach applies for Eqs.~\eqref{eq:back-x} and~\eqref{eq:back-w} in the backward pass. 

The moduli set $\mathcal{M}$ must be chosen to ensure that the outputs of the RNS operations are smaller than $M$, which means that
\begin{equation}
    \log_2 M \geq b_{\out} = b_{\inp} + b_w + \log_2(h)-1,
    \label{eq:rns_bit_inequality}
\end{equation}
should be guaranteed for a dot product between $b_{\inp}$-bit input and $b_w$-bit weight vectors with $h$-elements.
This constraint prevents overflow in the computation.

%% file: text/2-2-results-precision.tex
\subsection*{Precision and Energy Efficiency in the RNS-based Analog Core}
\label{sec:precision}

\input{text/fig-energy-eff}



The selection of moduli set $\mathcal{M}$, which is constrained by Eq.~\eqref{eq:rns_bit_inequality}, impacts the achievable precision at the output as well as the energy efficiency of the RNS-based analog core. 
Table~\ref{table:moduli-sets} compares RNS-based analog GEMM cores with example moduli sets and regular fixed-point analog GEMM cores. 
Here, we show two cases for the regular fixed-point representation: (1) the low-precision (LP) case where $b_{\out} > b_{\adc} = b_{\dac}$, and (2) the high-precision (HP) case where $b_{\out} = b_{\adc} > b_{\dac}$. 
It should be noted that all three analog cores represent data as fixed-point numbers. We use the term `regular fixed-point core' to refer to a typical analog core that performs computations in the standard representation (without RNS). `RNS-based core' refers to an analog core that performs computations on the fixed-point residues.

While the LP approach introduces $b_{\out}- b_{\adc}$ bits of information loss in every dot product, the HP approach uses high-precision ADCs to prevent this loss.
For the RNS-based core, we picked $b_{\inp} = b_{w} = b_{\adc} = b_{\dac} = \lceil \log_2m_i\rceil \equiv b$ for ease of comparison against the fixed-point cores.
Table~\ref{table:moduli-sets} shows example moduli sets that are chosen to guarantee Eq.~\eqref{eq:rns_bit_inequality} for $h = 128$ while keeping the moduli under the chosen bit-width $b$.
In this case, for $n$ moduli with bit-width of $b$, $M$ covers $\approx n \cdot b$ bits of range at the output.
$h$ is chosen to be 128 as an example considering the common layer sizes in the evaluated \textsf{MLPerf (Inference: Datacenter)} benchmarks. 
The chosen $h$ provides high throughput with high utilization of the GEMM core.

\begin{figure}[t]
\centering
\includegraphics[width=\linewidth]{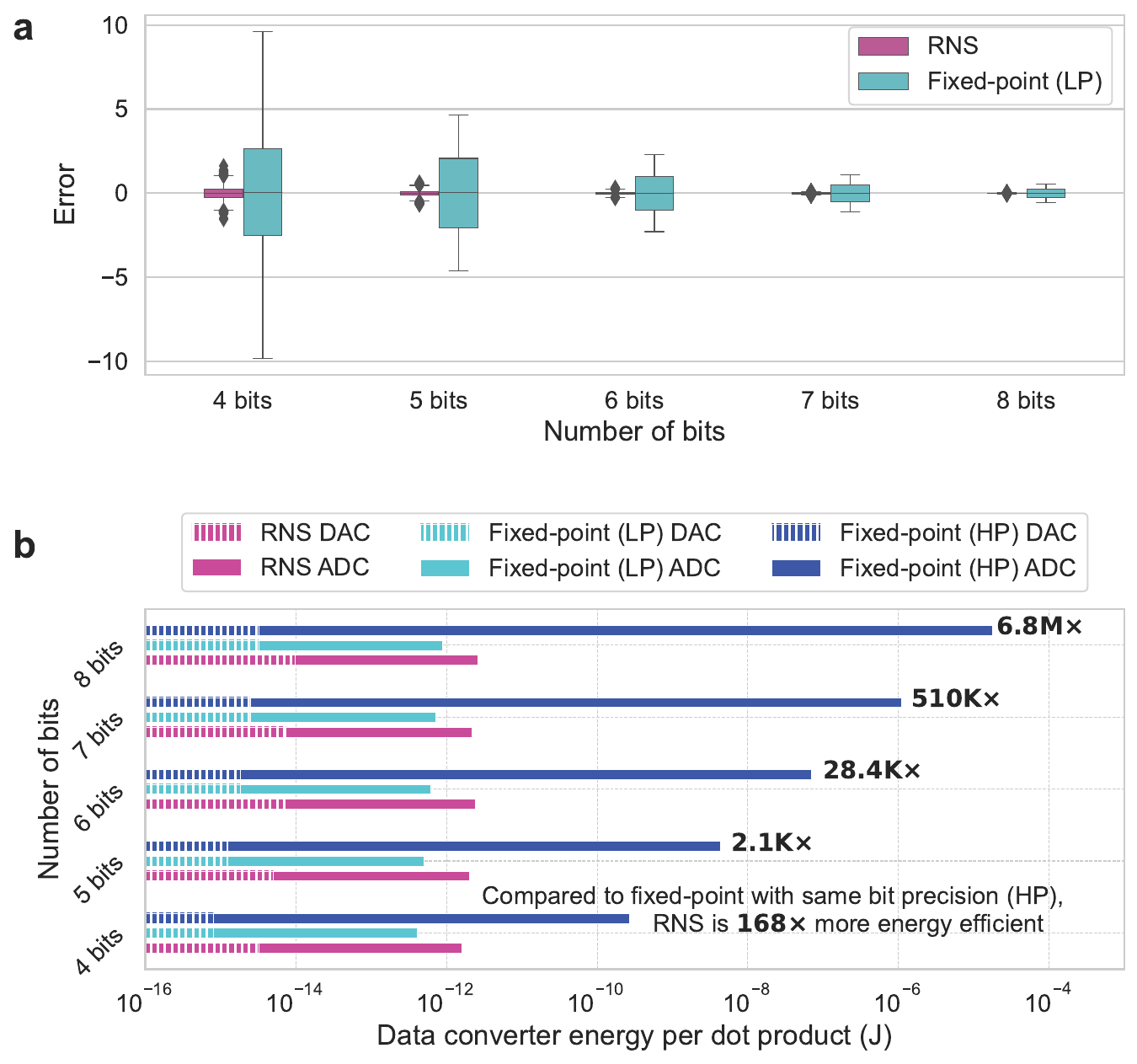}
\caption{\scriptsize{\textbf{Comparison of the RNS-based and regular fixed-point analog approaches.} \textbf{a} The distribution of average error observed at the output of a dot product performed with the RNS-based analog approach (pink) and the LP regular fixed-point analog approach (cyan). Error is defined as the distance from the result calculated in FP32. The experiments are repeated for 10,000 randomly generated vector pairs with vector size $h=128$. \textbf{b} Energy consumption of data converters (i.e., DACs and ADCs) per dot product for the RNS-based analog approach (pink) and the LP (cyan) and HP (dark blue) regular fixed-point analog approaches. See Methods for the energy estimation methodology. }}

\label{fig:rns-precision}
\end{figure}

Fig.~\ref{fig:rns-precision}a compares the error (with respect to FP32 results) observed when performing dot products with the RNS-based core and the LP fixed-point core with the same bit precision.
Both cores use the configurations described in Table~\ref{table:moduli-sets} for the example vector size $h = 128$.
The larger absolute error observed in the LP fixed-point case illustrates the effect of the information loss mentioned above.
HP fixed-point case is not shown as it is equivalent to the RNS case.


Fig.~\ref{fig:rns-precision}b shows the energy consumption of DACs and ADCs per dot product for the three aforementioned analog hardware configurations.
To achieve the same MVM throughput as the (LP/HP) fixed-point cores, the RNS-based core with $n$ moduli must use $n$ distinct MVM units and $n$ sets of DACs and ADCs.
This makes the energy consumption of the RNS-based core $n \times$ larger compared to the LP fixed-point approach. 
However, the LP fixed-point approach with low-precision ADCs experiences information loss in the partial outputs and hence has lower accuracy. 

The RNS-based approach and the HP fixed-point approach provide the same bit precision (i.e., the same DNN accuracy).
Yet, using the RNS-based approach is orders of magnitude more energy-efficient than the HP fixed-point approach. 
This is mainly because of the high cost of high-precision ADCs required to capture the full output in the HP fixed-point approach. 
ADCs dominate the energy consumption with approximately three orders of magnitude higher energy usage than DACs with the same bit precision.
In addition, energy consumption in ADCs increases exponentially with increasing bit precision~\cite{murmann-mixed-signal}.
This favors using multiple DACs and ADCs with lower precision in the RNS-based approach over using a single high-precision ADC.
Briefly, the RNS-based approach provides a sweet spot between LP and HP fixed-point approaches without compromising on both high accuracy and high energy efficiency.

\begin{table*}[t]
    \centering
    
    \caption{\scriptsize{Data and data converter precision in RNS-based, LP fixed-point, and HP fixed-point analog cores.} 
    }
    \label{table:moduli-sets}
    \begin{tabular}{ccccccccccccc}
        \toprule
        &\multicolumn{5}{c}{\textbf{RNS-based Core (This work)}} & \multicolumn{4}{c}{\textbf{LP Fixed-Point Core}} &\multicolumn{3}{c}{\textbf{HP Fixed-Point Core}} \\
        \cmidrule(lr){2-6}
        \cmidrule(lr){7-10}
        \cmidrule(lr){11-13}
        $b_{\inp}$, $b_w$ & $b_{\dac}$ & ${\log}_2\mathcal{M}$ & $b_{\adc}$ &  Moduli Set $(\mathcal{M})$ & RNS Range ($M$) & $b_{\dac}$ & $b_{\out}$ & $b_{\adc}$  & Lost Bits & $b_{\dac}$ & $b_{\out}$ & $b_{\adc}$\\ 
        \cmidrule(lr){1-1}
        \cmidrule(lr){2-2}
        \cmidrule(lr){3-3}
        \cmidrule(lr){4-4}
        \cmidrule(lr){5-5}
        \cmidrule(lr){6-6}
        \cmidrule(lr){7-7}
        \cmidrule(lr){8-8}
        \cmidrule(lr){9-9}  
        \cmidrule(lr){10-10}
        \cmidrule(lr){11-11}
        \cmidrule(lr){12-12}
        \cmidrule(lr){13-13}
        4 & 4 &4 & 4 & $\{15, 14, 13, 11\}$ & $\simeq 2^{15} - 1$ &4& 14 & 4 & 10&4& 14 & 14  \\
        5 & 5 & 5 &5 & $\{31, 29, 28, 27\}$ & $\simeq 2^{19} - 1$ &5& 16 & 5  & 11 &5& 16 & 16 \\
        6 & 6 & 6 &6 & $\{63, 62, 61, 59\}$ & $\simeq 2^{24} - 1$ &6& 18 & 6  & 12 &6& 18 & 18  \\
        7 & 7 & 7 &7 & $\{127, 126, 125\}$ & $\simeq 2^{21} - 1$ &7& 20 & 7 & 13 &7& 20 & 20 \\
        8 & 8 & 8 & 8 &$\{255, 254, 253\}$ & $\simeq 2^{24} - 1$ &8& 22 & 8 & 14 &8& 22 & 22 \\
        \bottomrule
    \end{tabular}
\end{table*}

\input{text/fig-no-err-acc}
\input{text/fig-dataflow}

%% file: text/fig-energy-eff.tex


%% file: text/fig-no-err-acc.tex
\begin{figure*}[ht]
\centering
\scriptsize
\includegraphics[width=\textwidth]{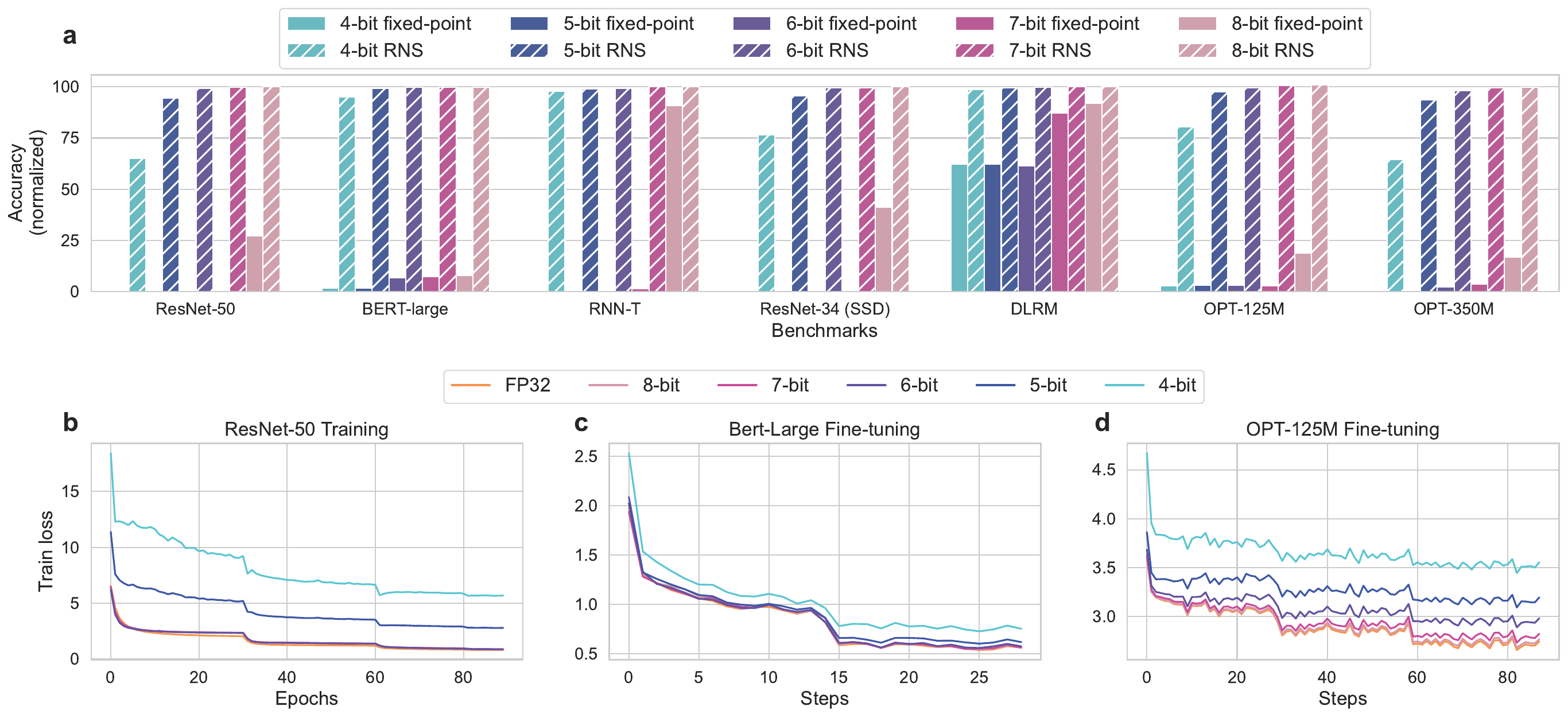}

\caption{\scriptsize{\textbf{Accuracy performance of the RNS-based analog core.} \textbf{a} Inference accuracy of regular fixed-point (LP) and RNS-based cores (See Table~\ref{table:moduli-sets}) on \textsf{MLPerf (Inference: Datacenters)} benchmarks. The accuracy numbers are normalized to the FP32 accuracy. 
\textbf{b-d} Loss during training for FP32 and RNS-based approaches with varying moduli bit-width.  ResNet50 \textbf{(a)} is trained from scratch for 90 epochs using SGD optimizer with a momentum. BERT-Large \textbf{(b)} and OPT-125M \textbf{(c)} are fine-tuned from pre-trained models. Both models are fine-tuned using the Adam optimizer with a linear learning rate scheduler for 2 and 3 epochs for BERT-Large and OPT-125M, respectively. 
All inference and training experiments use FP32 for all non-GEMM operations.
}}

\label{fig:rns-accuracy}
\end{figure*}

%% file: text/fig-dataflow.tex
\begin{figure*}[ht]
\centering
\includegraphics[width=.7\linewidth]{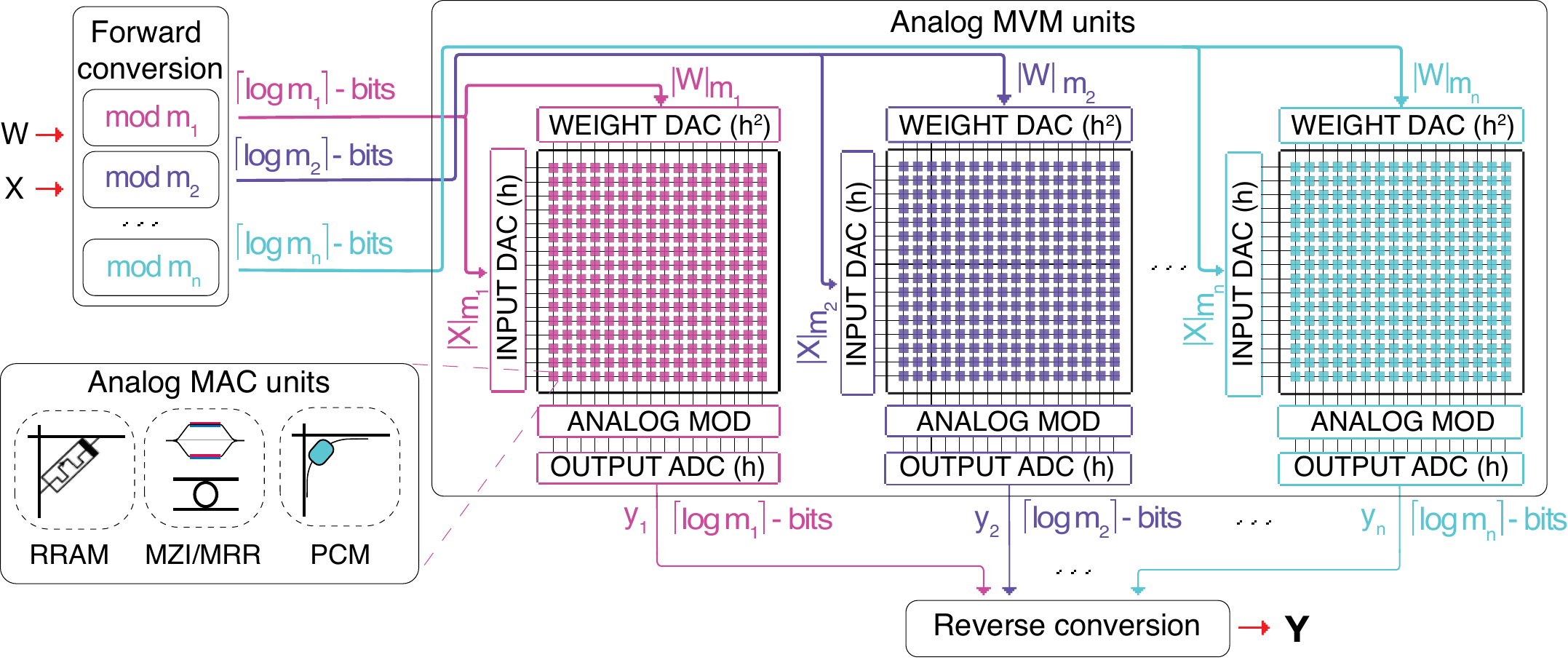}
\caption{\scriptsize{\textbf{RNS-based analog GEMM dataflow.} The operation is shown for a moduli set  $\mathcal{M} = \{m_1, \dots, m_{n}\}$. The $n$ $h\times h$ analog MVM units are represented as generic blocks. The dataflow is agnostic of the technology. }}
\label{fig:rns-core}
\end{figure*}

%% file: text/2-3-results-accuracy.tex


\subsection*{Accuracy in the RNS-based Analog Core}

Fig.~\ref{fig:rns-accuracy}a compares the inference accuracy of \textsf{MLPerf (Inference: Datacenters)} benchmarks~\cite{ml-perf-2020} and OPT~\cite{zhang2022opt} (a transformer-based LLM) when run on an RNS-based analog core and a fixed-point (LP) analog core.
The HP fixed-point analog core results are not shown as they are equivalent to the RNS-based results.
The evaluated DNNs, their corresponding tasks, and the datasets are listed in Table~\ref{table:dnns}.
Fig.~\ref{fig:rns-accuracy}a shows that the RNS-based approach significantly ameliorates the accuracy drop caused by the low-precision ADCs used in the LP fixed-point approach for all the networks.
By using the RNS-based approach, it is possible to achieve ${\geq}99\%$ of FP32 accuracy (this cut-off is defined in the MLPerf benchmarks~\cite{ml-perf-2020}) for all evaluated benchmarks when using residues with as low as $6$ bits.
This number can be lowered to $5$ bits for BERT-large and RNN-T and to $4$ bits for DLRM. 

Besides inference, the RNS-based approach opens the door for analog computing to be used in tasks that require higher precision than inference such as DNN training.
Figure~\ref{fig:rns-accuracy}b shows the loss during DNN training/fine-tuning.
Table~\ref{table:training} reports the validation accuracies after FP32 and RNS-based low-precision training.
Here, the GEMM operations during forward and backward passes of training follow the same methodology as inference, with weight updates carried out in FP32.
Our experiments show that ${\geq}99\%$ FP32 validation accuracy is achievable after training ResNet50 from scratch using the RNS-based approach with only $6$-bit moduli.
Similarly, fine-tuning BERT-large and OPT-125M by using $5$-bit and $7$-bit moduli, respectively, can reach ${\geq}99\%$ FP32 validation accuracy. 
The results are noticeably promising as the previous efforts on analog DNN hardware that adopted the LP fixed-point approach had never successfully demonstrated the training of state-of-the-art DNNs due to the limited precision of this approach.

Fig.~\ref{fig:rns-core} illustrates the dataflow of the RNS-based analog core when performing MVM as part of the DNN inference/training.
An input vector $X$ and a weight matrix $W$ to be multiplied in the MVM unit are first mapped to signed integers. 
To mitigate the quantization effects, $X$ and each row in $W$ are scaled by an FP32 scaling factor that is unique to the vector (See Methods). 
The signed integers are then converted into RNS residues through modulo operation (i.e., forward conversion).
By construction, each residue is within the range of $[0,m_i)$.
To achieve the same throughput as a fixed-point analog core, the RNS-based analog core with $n$ moduli requires using $n$ analog MVM units---one for each modulus---and running them in parallel.
Each analog MVM unit requires a set of DACs for converting the associated input and weight residues into the analog domain.
The MVM operations are followed by an analog modulo operation on each output residue vector.
Thanks to the modulo operation, the output residues---to be captured by ADCs---are reduced back to the $[0,m_i)$ range. 
Therefore, a bit precision of $\lceil\log_2{m_i}\rceil$ is adequate for both DACs and ADCs to perform input and output conversions without any information loss. 
The output residues are then converted back to the standard representation in the digital domain using Eq.~\eqref{eq:crt} to generate the signed-integer output vector, which is then mapped back to an FP32 final output $Y$.
The non-linear function $f$ (e.g., ReLU, sigmoid, etc.) is then performed digitally in FP32.

\begin{table}[t]
\caption{\scriptsize{\textsf{MLPerf (Inference: Datacenters)} benchmarks. }}
\label{table:dnns}
\begin{tabular}{  m{2.4cm}  m{3.2cm} m{2.6cm} } 
\toprule
\textbf{DNN } & \textbf{Task}        & \textbf{Dataset}\\ 
\cmidrule(lr){1-1}
\cmidrule(lr){2-2}
\cmidrule(lr){3-3}
ResNet50   & Image classification & ImageNet~\cite{imagenet}       \\ 
SSD-ResNet34 & Object detection     & MS COCO~\cite{coco}       \\ 
BERT-Large    & Question answering   & SQuADv1.1~\cite{squad}      \\ 
RNN-T        & Speech recognition   & Librispeech~\cite{librispeech}     \\ 
DLRM         & Recommendation       & 1TB Click Logs~\cite{ijcai2021p290} \\ 
OPT-125M          & Language Modeling       & Wikitext~\cite{wikitext} \\ 
OPT-350M        & Language Modeling       & Wikitext \\ 
\bottomrule
\end{tabular}
\end{table}

\begin{table}[t]
\caption{\scriptsize{Validation accuracy results after training/fine-tuning.}  }
\begin{tabular}{p {1.5cm}>{\centering}p{1.7cm}>{\centering}p{2.3cm}>{\centering\arraybackslash}p{2.6cm}}
\toprule
 & \textbf{ResNet50} & \textbf{BERT-Large} & \textbf{OPT-125M}  \\

 \textbf{Precision}  & \textbf{Acc.($\%$)} & \textbf{F1 Score ($\%$)} & \textbf{Acc.($\%$)/PPL}  \\  
\cmidrule(lr){1-1}
\cmidrule(lr){2-2}
\cmidrule(lr){3-3}
\cmidrule(lr){4-4}
FP32 &75.80 & 91.03 & 43.95/19.72\\
8-bit &75.77 & 90.98 & 43.86/20.00 \\
7-bit &75.68 & 90.97 & 43.59/20.71 \\
6-bit &75.13 & 90.85 & 42.79/22.62\\
5-bit &59.72 & 90.81 & 41.45/26.17 \\
4-bit &42.15 & 89.66 & 38.64/35.65 \\
\bottomrule
\label{table:training}
\end{tabular}
\end{table}

%% file: text/2-4-results-error-corr.tex
\subsection*{Redundant RNS for Fault Tolerance}
\label{sec:rrns}

Analog compute cores are sensitive to noise.
In the case of RNS, even small errors in the residues can result in a large error in the corresponding integer they represent.
The Redundant Residue Number System (RRNS)~\cite{rrns-codes-hanzo, rrns-2015,yang2001redundant} can detect and correct errors---making the RNS-based analog core fault tolerant.
RRNS uses a total of $n+k$ moduli: $n$ non-redundant and $k$ redundant. 
An RRNS($n+k,n$) code can detect up to $k$ errors and can correct up to $\lfloor \frac{k}2 \rfloor$ errors.
In particular, the error in the codeword (i.e., the $n+k$ residues representing an integer in the RRNS space) can be one of the following cases:
\begin{itemize}
    \item \textbf{Case 1:} Fewer than $\lfloor \frac{k}2 \rfloor$ residues have errors---thereby they are correctable,
    \item \textbf{Case 2:} Between $\lfloor \frac{k}2 \rfloor$ and $k$ residues have errors or the codeword with more than  $k$ errors does not overlap with another codeword in the RRNS space---thereby the error is detectable,
    \item \textbf{Case 3:} More than $k$ residues have errors and the erroneous codeword overlaps with another codeword in the RRNS space---thereby the error goes undetected.
\end{itemize}

Errors are detected by using majority logic decoding wherein we divide the total $n+k$ output residues into ${n+k} \choose n$ groups with $n$ residues per group. 
One simple way of majority logic decoding in this context is to convert the residues in each group back to the standard representation via CRT to generate an output value for each group and compare the results of the ${n+k} \choose n$ groups. 
If more than $50\%$ of the groups have the same result in the standard representation, then the generated codeword is correct.
This corresponds to \textbf{Case 1}.
In contrast, not having a majority indicates that the generated codeword is erroneous and cannot be corrected. 
This corresponds to \textbf{Case 2}.
In this case, the detected errors can be eliminated by repeating the entire calculation. 
In \textbf{Case 3}, the erroneous codeword generated by the majority of the groups overlaps with another codeword. 
As a result, more than $50\%$ of the groups have the same incorrect result and the error goes undetected.
To optimize the hardware performance of this process, more efficient base-extension-based algorithms~\cite{base-ext} instead of CRT can be used for error detection.

The final error probability in an RRNS code depends on the percentage of the \emph{non-correctable} errors observed in the residues. 
This probability is influenced by the chosen moduli set and the number of error correction iterations (See Methods).
Let $p_c, p_d,$ and $p_u$ be the probabilities that Cases 1, 2, and 3 occur respectively when computing a single output.
Overall, $p_c + p_d + p_u = 1$.
For a single attempt (i.e., $R=1$), the probability of producing the incorrect output integer is $p_\text{err} (R=1) = 1 - p_c = p_u+p_d$.
Generally, it is possible to repeat the calculations $R$-times until no detectable error is found at the expense of increasing compute latency.
In this case, the probability of having an incorrect output after $R$ attempts of error correction is 
\begin{equation}
p_\text{err}(R) =  1 - p_c  \sum_{r=0}^{R-1} (p_{d})^r.
\label{eq:p_err}
\end{equation}
As the number of attempts increases, the output error probability decreases and converges to $\lim_{R\to\infty} p_\text{err}(R) = p_u/(p_u+p_c)$.
 
Fig.~\ref{fig:therotical-err} shows $p_\text{err}$ for different numbers of redundant moduli $(k)$, attempts $(R)$, and moduli sets with different bit-widths. 
Broadly, as the probability of a single residue error $p$ increases, the output error probability tends to $1$.
For a given number of attempts, increasing bit precision and the number of redundant moduli decreases $p_\text{err}$. 
For a fixed number of redundant moduli and a fixed number of bits per moduli, $p_\text{err}$ decreases as the number of attempts increases. 

Fig.~\ref{fig:exp-acc-err} investigates the impact of noise on the accuracy of two large and important MLPerf benchmarks---ResNet50 and BERT-Large---when using RRNS.
The two networks show similar behavior: adding extra moduli and increasing the number of attempts decrease $p_\text{err}$ at the same value of $p$. 
ResNet50 requires ${\sim} 3.9$ GigaMAC operations (GOp) for one inference on a single input image. 
For a $128\times128$ MVM unit, inferring an ImageNet image through the entire network involves computing ${\sim}29.4$M partial output elements. 
Therefore, we expect the transition point from an accurate network to an inaccurate network to occur at $p_\text{err}$ to be $\leq1/29.4$M = $3.4\times 10^{-8}$.
This $p_\text{err}$ transition point is $\leq1/358.6$M = $2.8\times 10^{-9}$ for BERT-Large.
Fig.~\ref{fig:exp-acc-err}, however, shows that the evaluated DNNs are more resilient to noise than expected: it is able to tolerate higher $p_\text{err}$ while maintaining good accuracy. 
The accuracy of ResNet50 only starts degrading (below $99\%$ FP32) when $p_{\text{err}} \approx 4.5 \times10^{-5}$ ($1000 \times$ higher than the estimated value) on average amongst the experiments shown in Figure~\ref{fig:exp-acc-err}. 
This transition probability is $p_{\text{err}} \;{\approx} \; 4\times10^{-4}$ for BERT-Large (on average $100,000 \times$ higher than the estimated value).
\input{text/figs-err}

%% file: text/figs-err.tex
\begin{figure*}[ht]
\centering
\includegraphics[width=\textwidth]{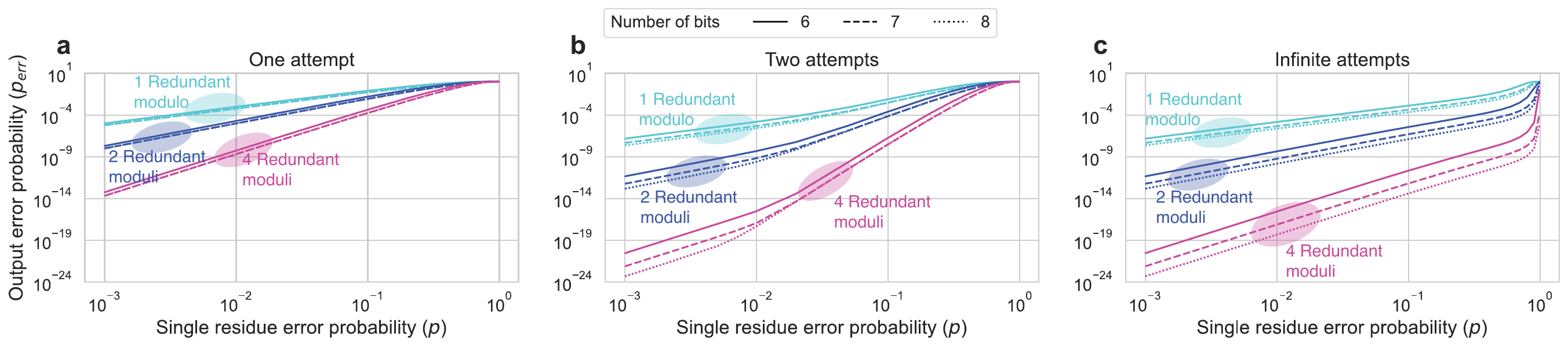}
\caption{\scriptsize{\textbf{Calculated output error probability ($\mathbf{p_{\text{err}}}$) versus single residue error probability ($\mathbf{p}$).} \textbf{a-c} $p_{\text{err}}$ for one (\textbf{a}), two (\textbf{b}), and infinite (\textbf{c}) error correction attempts and a varying number of redundant moduli $(k)$.}}

\label{fig:therotical-err}
\end{figure*}

\begin{figure*}[ht]
\centering
\includegraphics[width=\textwidth]{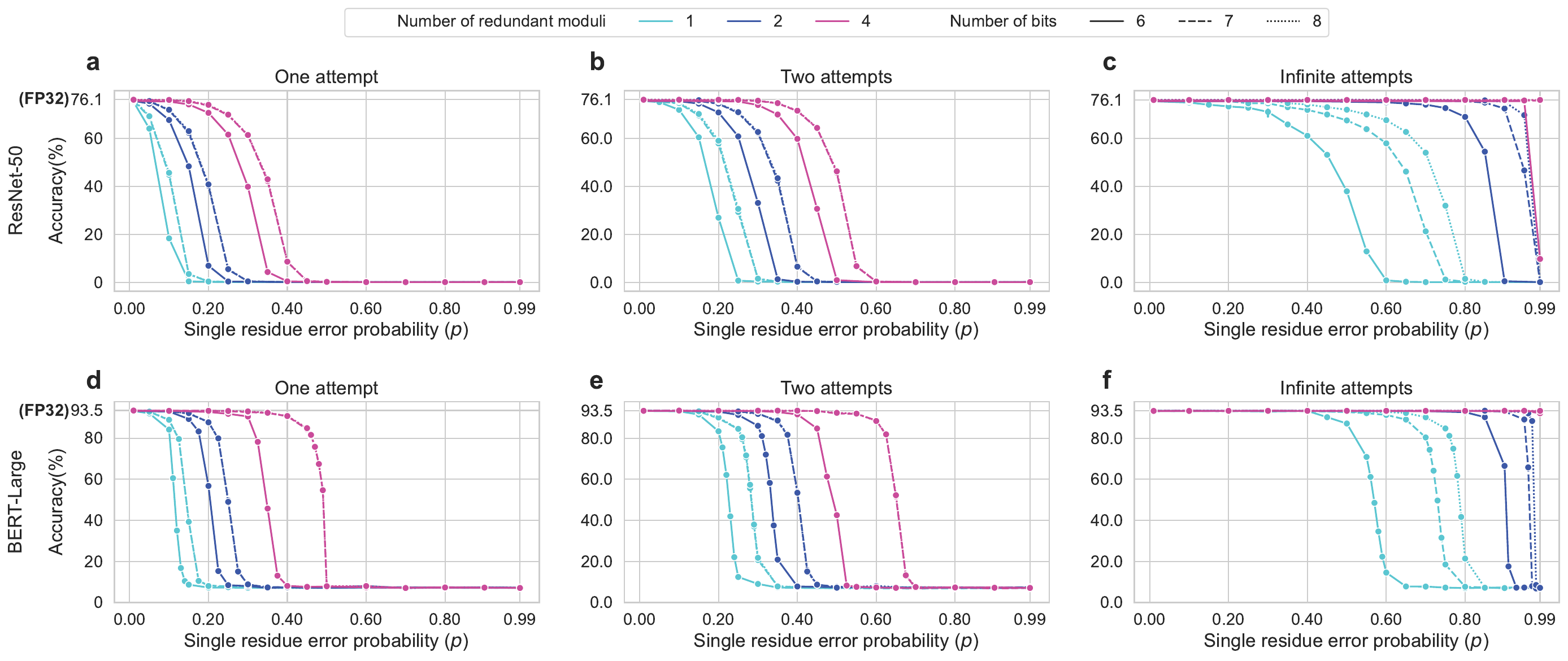}
\caption{\scriptsize{\textbf{Inference accuracy versus single residue error probability ($\mathbf{p}$).} \textbf{a-f} The plots show ResNet-50 (\textbf{a-c}) and BERT-Large (\textbf{d-f}) inference accuracy results under varying $p$ for one (\textbf{a} and \textbf{d}), two (\textbf{b} and \textbf{e}), and infinite (\textbf{c} and \textbf{f}) error correction attempts and a varying number of redundant moduli ($k$).}}

\label{fig:exp-acc-err}
\end{figure*}

%% file: text/3--discussion.tex
\section*{Discussion}


The RNS (and the fault-tolerant RRNS) framework are agnostic to the analog technology employed. 
Generally, the RNS GEMM operations can be performed as a regular GEMM operation followed by a modulo operation in the analog domain. 
Analog GEMM is well-explored in the literature. 
Previous works leveraged photonics~\cite{CoherentNanophotonic2017, 11-tops, weight-bank-2017, dnnara, pixel-2020, albireo-2021, adept}, crossbar arrays consisting of resistive RAM~\cite{yao2020fully, prime-2016, isaac, 1t1m-2016, tang-2017}, switched capacitors~\cite{bankman-2015, bankman-sc16}, PCM cells~\cite{feldmann2021parallel}, STT-RAM~\cite{jain2017computing, shi2020performance}, etc.

The analog modulo operation can be performed electrically or optically.
In the electronic domain, one can use ring oscillators: a circuit that generates a continuous waveform by cycling through a series of inverters \cite{analog-modulo-2018} to perform modulo operations.
By carefully designing the parameters of the ring oscillator, it is possible to create an output frequency that corresponds to the desired modulus value.
Alternatively, the phase of a signal can be used for performing modulo due to the periodicity of phases in optical systems. 
Optical phase is inherently modular against $2\pi$.
By modulating the phase of an optical signal, one can achieve modulo operations in the analog domain. 
Using RNS requires forward and reverse conversion circuits to switch between the RNS and the standard number system. 
The forward conversion is a modulo operation while the reverse conversion can be done using the CRT, mixed-radix conversion, or look-up tables. 
The (digital) hardware costs of these circuits can be reduced by choosing special moduli sets~\cite{hiasat1998residue, gallaher1997digit}. 

The RNS framework can be extended with the PNS to work with arbitrary precision, despite having DACs and ADCs with limited precision.
For applications requiring higher-precision arithmetic than the example cases in this study (e.g., some high-performance computing applications, homomorphic encryption, etc.), a higher $M$ value and therefore moduli with higher bit-width might be necessary, which will be bound by the same limitations discussed in this paper.
Instead, one can represent an integer value as $D$ separate digits where each digit is represented as a set of residues in the RNS domain and has an RNS range of $M$. 
This hybrid scheme can achieve $D \log_2M$ bit precision where $D$ can liberally be increased without increasing the bit precision of the data converters.
Different from the RNS-only scheme, the hybrid scheme requires overflow detection and carry propagation from lower digits to higher digits. 
The overflow detection can be achieved using two sets of residues: primary and secondary. 
While the operations are performed with both sets of residues, base extension between two sets helps detect any overflow and propagate the carry to the higher digits if required (See Methods). 


In conclusion, our work provides a methodology for precise, energy-efficient, and fault-tolerant analog DNN acceleration.
Overall, we believe that RNS is a crucial numeral system for the development of next-generation analog hardware capable of both inference and training state-of-the-art neural networks for advanced applications, such as generative artificial intelligence.

%% file: text/4--methods.tex
\section*{Methods}
\label{sec:methods}
\scriptsize
\subsection*{Handling Negative Numbers with RNS}
An RNS with dynamic range $M$ allows representing values within the range of $[0,M)$. 
This range can be shifted to $[-\psi,\psi]$, where $\psi = \lfloor (M-1)/2\rfloor$, to represent negative values.  
This is achieved by reassigning the values in between $(0, \psi]$ to be positive, 0 to be zero, and the numbers in between $(\psi,2\psi]$ to be negative (i.e. $[-\psi,0)$). 
Then, the values can be recovered uniquely by using CRT with a slight modification:
\begin{equation}
    A = 
    \begin{dcases}
        \sum_{i=1}^n|a_i M_i T_i|_M,& \text{if } \sum_{i=1}^n|a_i M_i T_i|_M \leq \psi\\
        \sum_{i=1}^n|a_i M_i T_i|_M - M,              & \text{otherwise}.
    \end{dcases}
     \label{eq:neg_crt}
\end{equation}

\subsection*{Data Converter Energy Estimation}
The DAC and ADC energy numbers in Fig.~\ref{fig:rns-precision} \textbf{(b)} are estimated by using equations formulated by Murmann~\cite{murmann-mixed-signal, murmann-adc-survey}. 
The energy consumption of a DAC per conversion is
\begin{equation}
        E_{\dac} = \text{ENOB}^2 C_u V_{\text{DD}}^2,
\label{eq:dac}
\end{equation}
where $C_u = 0.5$ fF is a typical unit capacitance and $V_{\text{DD}}$ = 1V is the supply voltage~\cite{murmann-mixed-signal}. The energy consumption of an ADC per conversion is estimated as
\begin{equation}
        E_{\adc} = k_1\text{ENOB} + k_24^\text{ENOB},
\label{eq:adc}
\end{equation}
where $k_1 {\approx}100$ fJ and $k_2 {\approx} 1$ aJ.
$E_{\adc}$ is dominated by the exponential term (i.e., $k_2 4^\text{ENOB}$) at large ENOB ( $\geq10$-bits).

\subsection*{Accuracy Modeling}

Both RNS-based and regular fixed-point analog cores are modeled using PyTorch for estimating inference and training accuracy. 
Convolution, linear, and batched matrix multiplication (BMM) layers are performed as tiled-GEMM operations which are computed tile-by-tile as a set of tiled-MVM operations.
Each input, weight, and output of the tiled MVM are quantized with a desired bit precision. 

Pre-quantization, the input vectors and weight tiles are first dynamically scaled, i.e. scaled at runtime, to mitigate the quantization effects as follows: 
For an $h\times h$ weight tile $\mathcal W_t$, we denote each row vector as ${\mathcal W}_{rt}$ where the subscript $r$ stands for the row and $t$ for the tile. 
Similarly, an input vector of length $h$ is denoted as $\mathcal X_t$ where $t$ indicates the tile. 
Each weight row $\mathcal W_{rt}$ shares a single FP32 scale $s_{rt}^w = \max (|\mathcal {W}_{rt}|)$ and each input vector $\mathcal X_t$ shares a single FP32 scale $s_{t}^x = \max (|\mathcal X_{t}|)$.
$h$ scales per $h\times h$ weight tile and 1 scale per input vector, in total $h+1$ scales, are stored for each tiled-MVM operation.
The tiled MVM is performed between the scaled weight and input vectors, $\widehat {\mathcal{W}}_{rt} = \mathcal W_{rt}/ s_{rt}^w$ and $\widehat {\mathcal X}_{t} = \mathcal X_{t}/ s_{t}^x$, to produce $\widehat Y_{rt} = \widehat {\mathcal{W}}_{t}\widehat {\mathcal X}_{t}$. 
The output $\widehat Y_{rt}$ is then quantized (if required) to resemble the output ADCs and multiplied back with the appropriate scales so that the actual output elements $Y_{rt} = \widehat Y_{rt} \cdot s_{rt}^w \cdot s_{t}^x$ are obtained.

Here, the methodology is the same for RNS-based and regular fixed-point cores. 
For the RNS-based case, in addition to the description above, the quantized input and weight integers are converted into the RNS space before the tiled-MVM operations. 
MVMs are performed separately for each set of residues and are followed by a modulo operation before the quantization step. 
The output residues for each tiled MVM are converted back to the standard representation using the CRT.

The GEMM operations (i.e., convolution, linear, and BMM layers) are sandwiched between an input operation $O_\text{in}$ and an output operation $O_\text{out}$.
This makes the operation order $O_\text{in}$-GEMM-$O_\text{out}$ during the forward pass, and  $O_\text{out}$-GEMM-$O_\text{in}$ in the backward pass. 
$O_\text{in}$ quantizes the input and weight tensors in the forward pass and is a null operation in the backward pass. 
In contrast, $O_\text{out}$ is a null operation in the forward pass and quantizes the activation gradients in the backward pass. 
In this way, the quantization is always performed before the GEMM operation. 
The optimizer (i.e., SGD or Adam) is modified to keep a copy of the FP32 weights to use during the weight updates. 
Before each forward pass, the FP32 weights are copied and stored. 
After the forward pass, the quantized model weights are replaced by the previously stored FP32 weights before the step function so that the weight updates are performed in FP32. 
After the weight update, the model parameters are quantized again for the next forward pass. 
This high-precision weight update step is crucial for achieving high accuracy in training.

In Fig.~\ref{fig:rns-accuracy}b, 
all the convolution, linear, and BMM layers in the models were replaced by the quantized versions.
We trained ResNet50 from scratch by using SGD optimizer for 90 epochs with a momentum of 0.9 and a learning rate starting from 0.1.
The learning rate was scaled down by 10 at epochs 30, 60, and 80.
We fine-tuned BERT-large and OPT-125M from the implementations available in the Huggingface transformers repository~\cite{Huggingface}.
We used the Adam optimizer for both models with the default settings. 
The script uses a linear learning rate scheduler.
The learning rate starts at 3e-05 and 5e-05 and the models are trained for 2 and 3 epochs, respectively for BERT-Large and OPT-125M.

\subsection*{Error distribution in the RRNS code space}

For an RRNS$({n+k},n)$ with $n$ non-redundant moduli $(m_1, m_2, ..., m_{n})$ and $k$ redundant moduli $(m_{n+1}, m_{n+2}, ..., m_{n+k})$, the probability distributions ($p_c$, $p_d$, and $p_u$) of different types of errors (Case 1, Case 2, and Case 3 that were mentioned in Redundant RNS for Fault Tolerance) are related to the Hamming distance distribution of the RRNS code space.
In an RRNS$({n+k},n)$, every integer is represented as ${n+k}$ residues ($r_i$ where $i\in \{1, ..., {n+k}\}$) and this vector of ${n+k}$ residues is considered as an RRNS codeword. 
A Hamming distance of $\eta \in \{0, 1, ..., {n+k}\}$ between the original codeword and the erroneous codeword indicates that $\eta$ out of ${n+k}$ residues are erroneous.
The erroneous codewords create a new vector space of ${n+k}$-long vectors where at least one $r_i$ is replaced  with $r_i'\neq r_i$ with $i\in \{1, ..., {n+k}\}$ and $r_i'<m_i$.
This vector space includes all the RRNS$({n+k},n)$ codewords as well as other possible ${n+k}$-long vectors that do not overlap with any codeword in the RRNS  code space. 
A vector represents a codeword and is in the RRNS code space if and only if it can be converted into a value within the legitimate range $[0,M)$ of the RRNS$({n+k},n)$ by using the CRT.
The number of all vectors that have a Hamming distance $\eta$ from a codeword in RRNS$({n+k},n)$ can be expressed as 
\begin{equation}
    V_\eta = \sum_{Q\binom{n+k}{\eta}}\prod_{i=1}^\eta (m_{i}-1),
     \label{eq:vector-distance}
\end{equation}
where ${Q\binom{n+k}{\eta}}$ represents one selection of $\eta$ moduli from ${n+k}$ moduli while $\sum_{Q\binom{n+k}{\eta}}$ represents the summation over all distinct $\binom{n+k}{\eta}$ selections.
The number of codewords that are in the RNS code space with a Hamming distance of $\eta \in \{ 0, 1, ..., {n+k}\}$ can be expressed as

\begin{equation}
    D_\eta = \sum_{h=0}^{\eta-1-k}(-1)^h \binom{n+k-\eta+h}{n+k-\eta} \zeta(n+k, \eta-h),
     \label{eq:code-distance}
\end{equation}
for $k+1\leq \eta\leq n+k$.
For $1\leq \eta\leq k$, $D_\eta=0$ and $D_0 = 1$. $\zeta(n+k, \eta)$ represents the total number of non-zero common divisors in the legitimate range $[0,M)$ for any $n+k-\eta$ moduli out of the $n+k$ moduli of the RRNS$(n+k, n)$ code and can be denoted as

\begin{equation}
    \zeta(n+k, \eta) =  \sum_{Q\binom{n+k}{n+k-\eta}}\Bigg\lfloor \frac{M-1}{m_{i_1}m_{i_2}...m_{i_{n+k-\eta)}}}\Bigg\rfloor,
     \label{eq:common-divisor}
\end{equation}
where $(m_{i_1},m_{i_2},...,m_{i_\lambda})$ with $1\leq \lambda \leq n+k$ is a subset of the RRNS$(n+k,n)$ moduli set. 

An undetectable error occurs only if a codeword with errors overlaps with another codeword in the same RRNS space. 
Given the distance distributions for the vector space $V$ and the codespace $D$ (Eq. \eqref{eq:vector-distance}, \eqref{eq:code-distance}, respectively), the probability of observing an undetectable error ($p_u$) for RRNS$(n+k,n)$ can be computed as

\begin{equation}
    p_{u} = \sum_{\eta=k+1}^{n+k}\frac{D_\eta}{V_\eta}p_E(\eta),
     \label{eq:p-undet}
\end{equation}
where $p_E(\eta)$ is the probability of having $\eta$ erroneous residues in a codeword which can be calculated as 
\begin{equation}
    p_{E}(\eta) = \sum_{Q\binom{n+k}{\eta}}p^\eta(1-p)^{(n+k-\eta)},
     \label{eq:p-error}
\end{equation}
for an error probability of $p$ in a single residue.

Eq.~\eqref{eq:code-distance} indicates that for up to $\eta=k$ erroneous residues $D_\eta = 0$, and so it is not possible for an erroneous codeword to overlap with another codeword in the RRNS code space. 
This guarantees the successful detection of the observed error. 
If the Hamming distance of the erroneous codeword is $\eta\leq\lfloor \frac{k}2 \rfloor$, the error can be corrected by the majority logic decoding mechanism.
In other words, the probability of observing a correctable error is equal to observing less or equal to $\lfloor \frac{k}2 \rfloor$ errors in the residues and can be calculated as 
\begin{equation}
    p_c = {\sum_{\eta=0}^{\lfloor \frac{k}2 \rfloor}}p_{E}(\eta) =\sum_{\eta=0}^{\lfloor \frac{k}2 \rfloor} \Big (\sum_{Q\binom{n+k}{\eta}}p^\eta(1-p)^{(n+k-\eta)}\Big ).
     \label{eq:p-correct}
\end{equation}
All the errors that do not fall under the undetectable or correctable categories are referred to as detectable but not correctable errors with a probability $p_d$ where $p_d =  1- (p_c + p_d)$.
The equations in this section were obtained from the work conducted by Yang~\cite{yang2001redundant}.

To model the error in the RNS core for the analysis shown in Fig.~\ref{fig:exp-acc-err},  $p_c$, $p_d$, and $p_u$ are computed for a given RRNS$(n+k,n)$ using Eqs.~\eqref{eq:p-undet} and~\eqref{eq:p-correct}. 
Given the number of error correction attempts, output error probability ($p_{err}$) is calculated according to Eq.~\eqref{eq:p_err}.
Random noise is injected at the output of every tiled-MVM operation using a Bernoulli distribution with the probability of $p_{err}$.

\subsection*{RNS Operations}
The proposed analog RNS-based approach requires modular arithmetic. 
In this section, we discuss two ways of performing modular arithmetic in the analog domain, one electrical and one optical.

\subsubsection*{Modular Arithmetic with Ring Oscillators}

In a ring oscillator, where each inverter has a propagation delay $t_\text{prop}>0$, there is always only one inverter that has the same input and output---either 1-1 or 0-0---at any given time when the ring oscillator is on. 
The location of this inverter with the same input and output propagates along with the signal every $t_\text{prop}$ time and rotates due to the ring structure.
This rotation forms a modular behavior in the ring when the location of this inverter is tracked.

Let $S_\text{RO}(t)$ be the state of a ring oscillator with $N$ inverters. Here, $S_\text{RO}(t) \in \{0, ..., N-1\}$ and $S_\text{RO}(t)=k$ means that the $k+1$-th inverter's input and output have the same value at time $t$.
$S_\text{RO}(t)$ keeps rotating between $0$ to $N-1$ as long as the oscillator is on. 
Figure~\ref{fig:rns-mod}a shows a simple example where $N=3$.
In the first $t_\text{prop}$ time interval, the input and output of the first inverter are both 0, therefore, the state $S_\text{RO}(t<t_\text{prop}) = 0$. 
Similarly, when $t_\text{prop}<t<2t_\text{prop}$, the input and output of the second inverter are 1, so $S_\text{RO}(t_\text{prop}<t<2t_\text{prop}) = 1$.
Here, the time between two states following one another (i.e., $t_\text{prop}$) is fixed and $S_\text{RO}(t)$ rotates ($0,1,2,0,1,...$). 
Assume the state of the ring oscillator is sampled periodically with a sampling period of $T_s = A\cdot t_\text{prop}$. 
Then, the observed change in the state of the ring oscillator between two samples ($S_\text{RO}({t=T_s})-S_\text{RO}({t=0})$) is equivalent to $|A|_N$ where $A$ is a positive integer value.
Therefore, to perform a modulo against a modulus value $m$, the number of inverters $N$ should be equal to $m$.
The dividend number $A$ and the sampling period can be adjusted by changing the analog input voltage to a voltage-to-time converter (VTC).

In a modular dot product or an MVM operation, the dividend $A$ is replaced by the output of the dot product. 
Analog dot products can be performed using traditional methods with no change with any desired analog technology where output can be represented as an analog electrical signal (e.g., current or voltage) before the analog modulo.

\subsubsection*{Modular Arithmetic with Phase Shifters}
\begin{figure*}[t]
\centering
\includegraphics[width=0.95\linewidth]{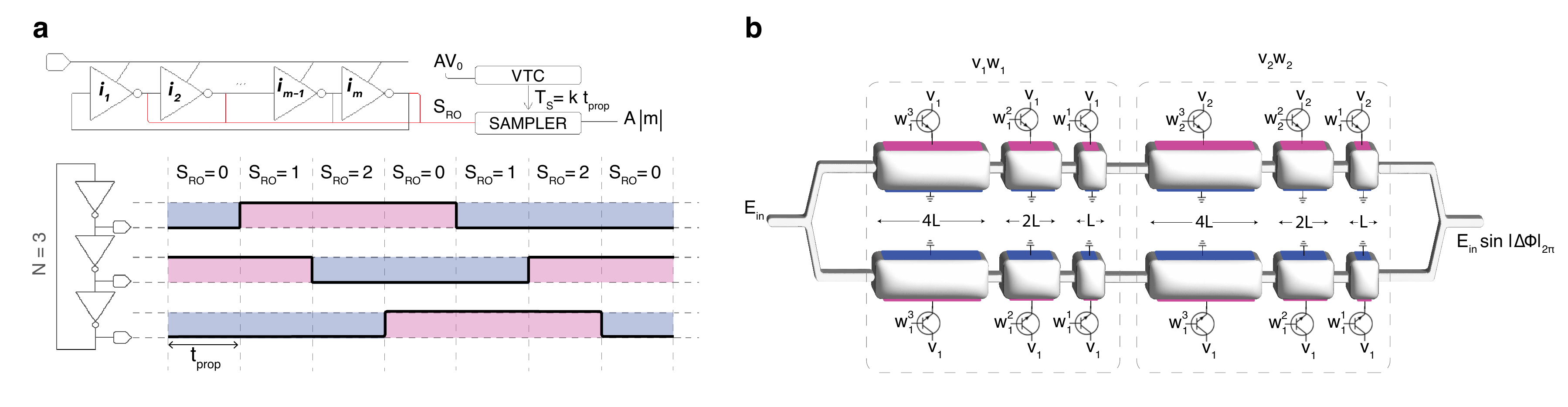}
\caption{\scriptsize{\textbf{Analog modulo implementations}. \textbf{a} Modulo operation performed using a ring oscillator. A ring oscillator with $N=3$ inverters is shown to perform modulo against a modulus $m = 3$. This operation is performed after every analog dot product to perform a modular dot product.} \textbf{b} Modular dot product performed using phase shifters. A modular dot product operation between two 2-element vectors $x$ and $w$, each with 3 digits, is shown by using a dual-rail set of cascaded phase shifters. The transistor switch turns on and supplies voltage to the phase shifter when the corresponding digit of $w$ is 1.}
\label{fig:rns-mod}
\end{figure*}
The amount of phase shift introduced by a single dual-rail phase shifter when $v$ and $-v$ voltages are applied on the upper and the bottom arms, respectively, is
\begin{equation}
    \Delta \Phi =  \frac{vL }{V_{\pi\cdot \text{cm}}},
\end{equation}
where $V_{\pi\cdot \text{cm}}$ is the modulation efficiency of the phase shifter and is a constant value. 
$\Delta \Phi$ is proportional to both the length of the shifter $L$ and the amount of applied voltage $v$. 
Figure~\ref{fig:rns-mod}b shows an example modular dot product operation between two vectors, $x$ and $w$, using cascaded dual-rail phase shifters. 
$w$ is encoded digit-by-digit using phase shifters with lengths proportional to $2^{j}$ where $j$ represents the digit number. 
In the example, each element (i.e., $w_0$ and $w_1$) of the 2-element vector $w$ consists of $3$ digits and uses $3$ phase shifters, each with lengths $L, 2L, $ and $4L$. 
If the $j$-th digit of the $i$-th element of $w$, $w_i^j = 1$, a voltage $v_i$ is applied to the phase shifter pair (top and bottom) with the length $2^{j}L$. 
If the digit $w_i^j=0$, then no voltage is applied and therefore no phase shift is introduced to the input signal. 
To encode the second operand $x$, a voltage $v_i$ that is proportional to $x_i$ is applied to all non-zero digits of $w_i$. 
To take modulo with a modulus $m$ instead of $2\pi$, the input $x$ and therefore the applied voltage $v$ should be multiplied by the constant $2\pi/m$. 
For encoding $x_i$,
\begin{equation}
    v_i = x_i \cdot \frac{V_{\pi\cdot \text{cm}}}{\pi L} \cdot \frac{2\pi}{m},
\end{equation}
should be applied so that the total phase shift at the end of the optical path is
\begin{equation}
    \Delta \Phi_{\text{total}} = \big| \frac{2\pi}{m} \sum_i\big(\sum_j (2^j w_i^j) x_i\big)\big|_{2\pi} = \frac{2\pi}{m} \big|\sum_i (w_ix_i)\big|_{m}.
\end{equation}

The resulting output values are collected at the end of the optical path and are in the form of the phase difference between input and output.  These outputs are then re-multiplied by $m/2\pi$ to obtain the outputs of the modular dot products for each residue.
 
\subsection*{Extended RNS}

By combining RNS and PNS, an integer value $Z$ can be represented as $D$ separate digits, $z_d$ where $d \in \{0, 1, ..., {D-1}\}$ and $0\leq z_d<M$ where $M$ is the RNS range:
\begin{equation}
    Z =  \sum_{d=0}^{D-1}z_dM,
\end{equation}
and can provide up to $D \log_2M$ bit precision. 
This hybrid scheme requires carry propagation from lower digits to higher digits, unlike the RNS-only scheme. 
For this purpose, one can use two sets of moduli, primary and secondary, where every operation is performed for both sets of residues. 
After every operation, overflow is detected for each digit and carried over to the higher-order digits. 

Let us define and pick $n_p$ primary moduli $m_i$, where $i\in \{1, ..., n_p\}$ and $n_s$ secondary moduli $m_j$, where $j\in \{1, ..., n_s\}$ and $m_i \ne m_j\ \forall \ \{i,j\}$.
Here $M = M_p \cdot M_s = \prod_{i=1}^{n_p}m_i \cdot \prod_{j=1}^{n_s}m_j$ is large enough to represent the largest possible output of the operations performed in this numeral representation and $M_p$ and $M_s$ are co-prime.

To execute an operation in this hybrid number system, the operation is performed separately for each digit of the output. 
These operations for each digit are independent of one another and can be parallelized except for the overflow detection and carry propagation. 
Assume $z_d= z_d|_{p;s}$ consists of primary and secondary residues and is a calculated output digit of an operation before overflow detection. 

$z_d$ can be decomposed as $z_d|_{p} = Q_d|_{p}M_p+R_d|_{p}$ where $Q_d|_{p}$ and $R_d|_{p}$ are the quotient and the remainder of the digit, with respect to the primary RNS. 
To detect a potential overflow in the digit $z_d$, a base extension from primary to secondary RNS is performed on $z_d|_{p}$ and the base extended residues are compared with the original secondary residues of the digit, $z_d|_{s}$. 
If the residues are the same, this indicates that there is no overflow, i.e., $Q_d|_{p;s}=0$, and both primary and secondary residues are kept without any carry moved to the next higher digit. 
As against that, if the base-extended secondary residues and the original secondary residues are not the same, it means that there exists an overflow (i.e., $Q_d|_{p;s}\ne 0$). 
In the case of overflow, the remainder of the secondary RNS, $R_d|_{s}$, is calculated through a base extension from primary to secondary RNS on $R_d|_{p}$ where $R_d|_{p} = z_d|_{p}$. 
$Q_d|_{s}$ can then be computed as $Q_d|_{s} = (z_d|_{s} - R_d|_{s}) M_p^{-1}$ where $|M_p \cdot M_p^{-1}|_{M_s} \equiv 1 $. 
$Q_d|_{p}$ is calculated through base extension from the secondary to primary RNS on the computed $Q_d|_{s}$. 
The full quotient $Q_d|_{p;s}$ is then propagated to the higher-order digit. 
Algorithm~\ref{alg:pseudocode} shows the pseudo-code for handling an operation $\square$ using the extended RNS representation.
The operation can be replaced by any operation that is closed under RNS. 

It should be noted that $z_d|_{p;s}$ is not always computed as $x_d|_{p;s} \square y_d|_{p;s}$. 
For operations such as addition, each digit before carry propagation is computed by simply adding the same digits of the operands, i.e., $z_d|_{p;s} = x_d|_{p;s} + y_d|_{p;s}$.
However, for multiplication, each digit of $z_d|_{p;s}$ should be constructed as in long multiplication.
The multiplication of two numbers in the hybrid number system with $D_x$ and $D_y$ digits requires $D_x D_y$ digit-wise multiplications and the output will result in $D_z = D_x + D_y$ digits in total. 
Similarly, a dot product is a combination of multiply and add operations. 
If two vectors with $h$ elements where each element has $D_x$ and $D_y$ digits, the output will require in $D_z = D_x + D_y + \log_2h$ digits.



 \begin{algorithm}[H] 
 \caption{\scriptsize \textbf{Pseudocode for performing the operation $\square$ using the hybrid number system.} Here, $x$ and $y$ are the inputs for operation $\square$ and $z$ is the output with $D$ digits. $z_d$ represents the digits of the output where $z_d|_p$ are the primary residues and $z_d|_s$ are the secondary residues. Primary and secondary residues together are referred to as $z'_{d}|_{p;s}$. $Q$ is the quotient and $R$ is the remainder where $z_d = Q_dM_p+R_d$. $\text{p2s}()$ and $\text{s2p}()$ refer to base extension algorithms from primary to secondary residues and from secondary to primary residues, respectively. }
 
\label{alg:loop}
\scriptsize{
\begin{algorithmic}[1]
\State $Q_{-1}|_{p;s}=0$
\For{$d \gets 0$ to $D_z$} 
\State $z'_d|_{p;s}={(x|_{p;s}}\ \square \ {y|_{p;s})_d} $
\EndFor
\For{$d \gets 0$ to $D_z$} 
\State $z_{d}|_{p;s} = z'_{d}|_{p;s}+ Q_{d-1}|_{p;s}$
\State $R_d|_p =z_{d}|_{p} $
\State $R_d|_s = \text{p2s}(R_d|_p) $ 

\If {$R_d|_s = z'_{d}|_{s}$}
\State $Q_{d}|_{p;s} = 0$
\Else 
\State $Q_{d}|_{s} = (z'_{d}|_{s} - R_d|_s) M_p^{-1}$
\State  $Q_{d}|_{p} = \text{s2p}(Q_d|_s)$
\EndIf

\EndFor

\end{algorithmic}}
\end{algorithm}

%% file: acknowledgements.tex
\section*{Acknowledgements}
We thank Dr. Rashmi Agrawal and Prof. Vijay Janapa Reddi for their insightful discussions. 

%% file: contributions.tex
\section*{Author Contributions}

D.B. conceived the project idea. C.D. and D.B. developed the theory. C.D. implemented the accuracy modeling and the analytical error models with feedback from D.B. and A.J. C.D. and L.N. conducted the experiments. D.B. and A.J. supervised the project. C.D. wrote the manuscript with input from all authors.

%% file: competing_interests.tex
\section*{Competing Interests}

The authors declare the following patent application: U.S. Patent Application No.: 17 / 543,676. L.N. and D.B. declare individual ownership of shares in Lightmatter, a startup company developing photonic hardware for AI.

%% file: references.tex
{}